\documentclass[10pt]{article}
\usepackage{amssymb}
\begin{document}
\def\nmonth{\ifcase\month\ \or January\or
   February\or March\or
April\or May\or June\or July\or August\or
   September\or October\or
November\else December\fi}
\def\nmonth{\ifcase\month\ \or January\or
   February\or March\or April\or May\or June\or July\or August\or
   September\or October\or November\else December\fi}
\def\rightheadline{\hfill\folio\hfill}
\def\leftheadline{\hfill\folio\hfill}
\newtheorem{theorem}{Theorem}[section]
\newtheorem{lemma}[theorem]{Lemma}
\newtheorem{remark}[theorem]{Remark}
\def\operatorname#1{{\rm#1\,}}
\def\text#1{{\hbox{#1}}}
\def\qedbox{\hbox{$\rlap{$\sqcap$}\sqcup$}}
\def\BB{{\mathcal{B}_1}}
\def\CC{{\mathcal{B}_2}}
\def\BX{{\mathcal{B}_{DR}}}
\def\BD{{\mathcal{B}_D}}
\def\BR{{\mathcal{B}_R}}
\def\B{{\mathcal{B}}}
\def\tr{{\operatorname{Tr}}}
\def\dvol{{\operatorname{dvol}}}
\newcommand{\reals}{\mathbf{R}}
\newcommand{\nats}{\mbox{${\rm I\!N }$}}

\def\gat{\gamma_a^T}
\def\la{\lambda}
\def\om{\omega}
\def\La{\Lambda}
\def\Th{\theta}
\def\chs{\chi^\star}
\def\Dirac{i\partial\!\!\!\!/}
\def\dirac{i\partial\!\!\!\!/-A\!\!\!\!/}
\def\pip{\Pi_+}
\def\pim{\Pi_-}
\def\pipl{\Pi_+^\star}
\def\pimi{\Pi_-^\star}
\def\gf{\tilde{\gamma}}
\def\rand{\left|_{\partial M}\right. }
\def\ch{\cosh \Theta}
\def\sh{\sinh\Theta}
\def\nen{\Lambda \ch -2\om\sh}
\def\tr{\mbox{Tr}}

\newcommand{\gc}{\gamma _M^0}
\newcommand{\gu}{\gamma _M^1}
\newcommand{\gt}{\tilde{\gamma}_M}
\newcommand{\gce}{\gamma _0}
\newcommand{\gue}{\gamma _1}
\newcommand{\beq}{\begin{eqnarray}}
\newcommand{\eeq}{\end{eqnarray}}
\newcommand{\nn}{\nonumber}
\makeatletter
  \renewcommand{\theequation}{%
   \thesection.\arabic{equation}}
  \@addtoreset{equation}{section}
\makeatother
\title{Finite temperature properties of the Dirac operator under
local boundary conditions}
\author{C.G. Beneventano\thanks{Fellow of CONICET - E-mail:
gabriela@obelix.fisica.unlp.edu.ar}\,,
E.M. Santangelo\thanks{Member of CONICET - E-mail: mariel@obelix.fisica.unlp.edu.ar}\\
Departamento de F{\'\i}sica, Universidad Nacional de La Plata\\
C.C.67, 1900 La Plata, Argentina}\maketitle
\begin{abstract}

We study the finite temperature free energy and fermion number for
Dirac fields in a one-dimensional spatial segment, under two
different members of the family of local boundary conditions
defining a self-adjoint Euclidean Dirac operator in two
dimensions. For one of such boundary conditions, compatible with
the presence of a spectral asymmetry, we discuss in detail the
contribution of this part of the spectrum to the zeta-regularized determinant of
the Dirac operator and,
thus, to the finite temperature properties of the theory.\\[.3cm]
{\bf Subject Classification} 

\end{abstract}
\section{Introduction}\label{intr}

When the Euclidean Dirac operator is considered on
even-dimensional compact manifolds with boundary, its domain can
be determined through a family of local boundary conditions which
define a self-adjoint boundary problem \cite{wipf95-443-201} (the
particular case of two-dimensional manifolds was first studied in
\cite{hras84-245-118}). The whole family is characterized by a
real parameter $\Th$, which can be interpreted as an analytic
continuation of the well known $\Th$ parameter in gauge theories.
These boundary conditions can be considered to be the natural
counterpart in Euclidean space of the well known chiral bag
boundary conditions.

Recently, it was shown \cite{bgks} that the boundary problem so
defined is not only self-adjoint, but also strongly elliptic
\cite{gilk95b,seel69-91-963} in any even dimension. Also in
reference \cite{bgks}, the meromorphic properties of the
associated zeta function were determined for manifolds of the
product type. For particular non-product manifolds, heat kernel
coefficients and zeta functions were treated in
\cite{espo02-66-085014}. Anomalies were also studied recently in
\cite{mv}.

One salient characteristic of these local boundary conditions is
the generation of an asymmetry in the spectrum of the Dirac
operator. For the particular case of two-dimensional product
manifolds, such asymmetry was shown, in \cite{bene02-35-9343}, to
be determined by the asymmetry of the boundary spectrum.

\bigskip

The aim of this paper is twofold:

i. To discuss a physical application of Euclidean bag boundary
conditions in two dimensions, with emphasis on the effect of the
spectral asymmetry.

ii. To (partially) answer the question posed in \cite{dunne}, as
to whether the fermion number is modified by temperature in low
dimensional bags.

\bigskip

In section \ref{Sect1}, we present the zero-temperature problem of
Dirac fermions subject to local boundary conditions with arbitrary
values of $\Th$ in $1+1$-dimensional Minkowski space-time, and
evaluate the vacuum energy and fermion number (equivalently,
$U(1)$ charge). This model has recently been considered, in the
context of brane theory, as the fermionic sector of an $N=2$
supersymmetric sigma model, coupled to a magnetic field at the
ends of the string \cite{lugo}.

In section \ref{Sect2}, we determine the spectrum of the Euclidean
Dirac operator at finite temperature for two particular values of
$\Th$. With these spectra at hand, we perform, in section
\ref{Sect3}, the calculation of the free energy via zeta function
regularization.

Section \ref{Sect4} is devoted to the evaluation of the free
energy in the grand-canonical ensemble and the fermion number,
through the introduction of a chemical potential (The
finite-temperature fermion number for a different local boundary
condition, not leading to a self-adjoint operator, was treated in 
\cite{dfls}).

Finally, section \ref{Sect5} presents a discussion of the main
results.

\section{Definition of the problem in Minkowski space-time}\label{Sect1}

We will use the metric (-,+), and choose for the Dirac matrices
\beq \gc=i\sigma_1,\qquad\gu=-\sigma_2\qquad {\rm and}\qquad
\gt=\gc\gu=\sigma_3.\label{gammam}\eeq

The action, for Dirac fermions coupled to a background field is
given by \beq S_M= i\int d^2x\, \bar{\Psi}(\dirac) \Psi \,.\eeq

Let us first particularize to the free case. The Hamiltonian can
be determined from the classical equation of motion
$\Dirac\Psi=0$, by proposing $\Psi(x^0,x^1)=e^{-iEx^0}\psi(x^1)$,
with $0\leq x^1 \leq L$. Thus, one gets the Hamiltonian $H=i\gt
\partial_1$. Its eigenfunctions are of the form \beq
\psi(x^1)=\left(\begin{array}{c}
  A e^{-i E x^1} \\
   B e^{i E x^1} \\
\end{array}\right)\,. \eeq

The boundary conditions will be taken to be \beq \left.\frac12
(1-i\gc
 e^{i\gt {\Th}_{0,L}}) \psi\right\rfloor_{x^1=0,L}=0\,.\eeq

 Now, it is easy to see that the eigenfunctions of the Hamiltonian depend only on the
 difference $\Th=\Th_L -\Th_0$, since the overall phase can always be
 eliminated through a constant chiral transformation. We will thus
 consider
 \beq
 \nn \left.\frac12 (1-i\gc ) \psi \right\rfloor_{x^1=0} &=& 0\\
 \left.\frac12 (1-i\gc e^{i\gt {\Th}}) \psi \right\rfloor_{x^1=L} &=& 0
 \,.\eeq

 Once such boundary conditions are imposed, the eigenfunctions
 read
 \beq
\psi_n(x^1)=\left(\begin{array}{c}
  A_n e^{-i E_n x^1} \\
   -A_n e^{i E_n x^1} \\
\end{array}\right)\,, \eeq
where \beq E_n =\frac{n\pi}{L}+\frac{\Th}{2L}\qquad {\rm with}
\qquad n=-\infty,...,\infty\,.\eeq

To evaluate the vacuum properties at zero temperature, let us
first consider $0<\Th<2\pi$. Then, when defined through zeta
function regularization \cite{zeta}, the Casimir energy is given by  \cite{cas}\beq \nn
E_C(\Th) &=& -\left.\frac12 \sum_n
|E_n|^{-s}\right\rfloor_{s=-1}\\
\nn &=&-\frac{\pi}{2L}\left(\zeta_H
(-1,1-\frac{\Th}{2\pi})+\zeta_H (-1,\frac{\Th}{2\pi})\right)
\\&=&\frac{\pi}{4L}\left(\frac{\Th^2}{2\pi^2}-\frac{\Th}{\pi}+\frac13\right)\,,\eeq
where $\zeta_H(s,u)$ is the Hurwitz zeta function \cite{rusa}

In particular, $E_C(\pi)=-\frac{\pi}{24L}$. Note that, for
$-2\pi<\Th<0$, the replacement $\Th\rightarrow 2\pi+\Th$ must be
performed. For $\Th=0$ \beq
E_C(0)=-\frac{\pi}{L}\zeta_R(-1)=\frac{\pi}{12
L}\,,\label{ec0}\eeq where $\zeta_R$ is the Riemann zeta function.
So, the Casimir energy is continuous at $\Th =0$.

As for the vacuum expectation value of the fermion number ($U(1)$
charge), again for $0<\Th<2\pi$, one has \cite{ns}
\beq \nn N(\Th)
&=& -\left.\frac12 \left(\sum_{E_n > 0} |E_n|^{-s}-\sum_{E_n < 0}
|E_n|^{-s}\right)\right\rfloor_{s=0}\\
\nn &=&\frac12 \left(\zeta_H (0,1-\frac{\Th}{2\pi})-\zeta_H
(0,\frac{\Th}{2\pi})\right)
\\&=&\frac12 \left(\frac{\Th}{\pi}-1\right)\,.\label{nn}\eeq

Also in this case,  for $-2\pi<\Th<0$ the replacement
$\Th\rightarrow 2\pi+\Th$ must be performed. For $\Th=\pi$, one
has $N(\pi)=0$. But, at variance with the Casimir energy, the
fermion number is discontinuous at $\Th=0$. In fact, in this case,
apart from a symmetric nonvanishing spectrum, a zero mode of the
Hamiltonian appears, which is its own charge conjugate. As a
consequence, $N(0)=\pm \frac12$ \cite{ns}.

In what follows, we will concentrate on two values of $\Th$, i.e.,
$\Th=0$ and $\Th=\pi$, to study the effect of a nonvanishing
temperature on both vacuum quantities. In both cases, the
Euclidean Dirac operator will turn to be self-adjoint, as shown in
\cite{wipf95-443-201}. Note these two boundary conditions are the
ones corresponding to Ramond and Neveu-Schwarz strings
\cite{lugo}.

\bigskip

\section{ Finite temperature. Spectrum of the Dirac operator}\label{Sect2}

In order to study the effect of temperature, we go to Euclidean
space, with the metric (+,+). To this end, we take the Euclidean
gamma matrices to be $\gce=-i\gc=\sigma_1$, $\gue=-\gu=\sigma_2$.
Thus, the Euclidean action is \beq S_E=-iS_M=\int d^2x\,
\bar{\Psi}(\dirac) \Psi\,,\eeq where the Euclidean zero component
of the gauge potential ($A_0$) is related to the corresponding
Minkowski one by $A_0=-iA_0^{M}$.

We start by treating the free case; then, the partition function
in the canonical ensemble is given by (see, for example, \cite{actor} and references therein)\beq \log{Z}=\log\,
det(\Dirac _{BC})\,.\eeq

Here, $BC$ stands for antiperiodic boundary conditions in the
``time" direction ($0\leq x_0 \leq \beta$, with
$\beta=\frac{1}{T}$) and, in the ``space" direction ($0\leq x_1
\leq L$), \beq
 \nn \left.\frac12 (1+\gce ) \Psi \right\rfloor_{0} &=& 0\\
 \left.\frac12 (1\pm \gce) \Psi \right\rfloor_{L} &=& 0\,.\label{bc}\eeq

 In the last equation, the plus sign corresponds to the case $\Th
 =0$, while the minus sign corresponds to $\Th =\pi$.

 In order to evaluate the partition function in the zeta
 regularization  approach, we first determine the eigenfunctions,
 and the corresponding eigenvalues ($\omega$), of the Dirac operator
\beq \Dirac\, \Psi=\left(\matrix{
  0&i\partial_0+\partial_1\cr
  i\partial_0-\partial_1& 0 } \right)\left(\begin{array}{c}
    \varphi(x_0,x_1) \\
    \chi(x_0,x_1) \\
  \end{array}\right)=\omega\left(\begin{array}{c}
    \varphi(x_0,x_1) \\
    \chi(x_0,x_1) \\
  \end{array}\right)\,.\eeq

  To satisfy antiperiodic boundary conditions in the
  $x_0$ direction, we expand
  \beq
  \Psi(x_0,x_1)=\sum_{\lambda}e^{i\lambda x_0}\psi(x_1)\,,\eeq
  with
  \beq
  \lambda_l=(2l+1)\frac{\pi}{\beta},\qquad
  l=-\infty,...,\infty\,.\label{lambda}\eeq

  After doing so we have, for each
  $\lambda_l$,
  \beq \nn (-\lambda_l+\partial_1)\chi&=&\omega
\varphi \\(-\lambda_l-\partial_1)\varphi&=&\omega \chi
\,.\label{diffeq}\eeq

 \subsection{$\Th=0$}\label{Sect21}

 It is easy to see that, with this boundary condition, no zero
 mode appears. For $\omega \neq 0$ one has, from (\ref{diffeq}),
 \beq \nn &&\partial^2_1\varphi=-\kappa^2\varphi
 \\&&\chi=-\frac{1}{\omega}(\lambda_l+\partial_1)\varphi\,,\eeq
where $\kappa^2=\omega^2-\lambda_l^2$.

For $\kappa \neq 0$, one has for the eigenvalues \beq \omega_{n,l}
=\pm \sqrt{\left(\frac{n\pi}{L}\right)^2+{\lambda_l}^2} \,,\qquad
{\rm with}\qquad n=1,...,\infty\,.\label{symspec}\eeq This part of
the spectrum is symmetric. In the case $\kappa=0$ one has a set of
$x_1$-independent eigenfunctions, corresponding to
\beq\omega_l=\lambda_l\,.\label{asymspec}\eeq

It is to be noted that $\omega_l=-\lambda_l$ are not eigenvalues;
so, this last portion of the spectrum will be asymmetric or not,
depending on whether the boundary spectrum ($\{\lambda_l\}$) is
so. This was to be expected from our result in
\cite{bene02-35-9343}, where we proved that, in this case, the
contribution of each boundary to the asymmetry equals one half the
boundary asymmetry, and the contributions from both boundaries add
when the boundary conditions are the same at both of them. For
$\lambda_l$ as in (\ref{lambda}), the asymmetry vanishes. However,
the contribution of the spectral asymmetry will be shown to be
crucial when evaluating the finite-temperature fermion number,
which will be done in section \ref{Sect4}.

\subsection{$\Th=\pi$}\label{Sect22}

Also in this case, there are no zero modes.

At variance from the case $\Th=0$, no solution exists for
$\kappa=0$, and only the symmetric part of the spectrum appears.
The eigenvalues are given by \beq \omega_{n,l} =\pm
\sqrt{\left((n+\frac12)\frac{\pi}{L}\right)^2+{\lambda_l}^2}
\,,\qquad {\rm with}\qquad n=0,...,\infty\,.\eeq

The absence of a nonsymmetric spectrum is also easy to understand
from our result in \cite{bene02-35-9343}, where it was shown that
the sign of the boundary contribution to the asymmetry changes
under a change of the intermediate sign in the projector defining
the boundary conditions. So, in this case, the contributions from
both boundaries cancel each other.

\section{Free energy}\label{Sect3}

With the eigenvalues of the Euclidean Dirac operator at hand, we
can now obtain the partition function, which is

\beq \log{Z}=\log\, det(\Dirac _{BC})\,.\eeq

When the determinant is defined through a zeta function
regularization \cite{zeta}, one has \beq
\left.\log{Z}=-\frac{d}{ds}\right\rfloor_{s=0}\,\zeta(s,\frac{\Dirac_{\,BC}}{\alpha})
\,.\label{partfunc}\eeq 

Here, $\zeta(s,\frac{\Dirac_{\,BC}}{\alpha})$ is the zeta function of the operator $\frac{\Dirac_{\,BC}}{\alpha}$ \cite{klaus} and, 
as usual, $\alpha$ is a parameter with
dimensions of mass, introduced to render the zeta function
dimensionless.

\subsection{$\Th=0$}\label{Sect31}

We will first discuss in detail the case where the boundary
conditions are determined by $\Th=0$. We will then have two types
of contributions. The first one comes from the symmetric part of
the spectrum, equation (\ref{symspec}); the second, from the
``nonsymmetric" part, equation (\ref{asymspec}). These two
contributions are given by

\beq \Delta_1=\left. -\frac{d}{ds}\right\rfloor_{s=0}{\zeta}_1
(s)\,,\label{del1}\eeq where \beq {\zeta}_1
(s)=(1+(-1)^{-s})\sum_{\begin{array}{c}
  n=1 \\
  l=-\infty \\
\end{array}}^{\infty}\left[{\left(\frac{n\pi}{\alpha L}\right)}^2+
{\left((2l+1)\frac{\pi}{\alpha
\beta}\right)}^2\right]^{-\frac{s}{2}}\,,\label{z1}\eeq and \beq
\Delta_2=\left. -\frac{d}{ds}\right\rfloor_{s=0}{\zeta}_2
(s)\,,\label{del2}\eeq where \beq \nn {\zeta}_2
(s)&=&\sum_{l=-\infty }^{\infty}\left[ (2l+1)\frac{\pi}{\alpha
\beta}\right]^{-s}\\&=&(1+(-1)^{-s})\sum_{l=0 }^{\infty}\left[
(2l+1)\frac{\pi}{\alpha \beta}\right]^{-s}\,.\label{z2}\eeq

Note that, in equations (\ref{z1}) and (\ref{z2}), $(-1)^{-s}$ is
undetermined (except at $s=0$), since it can be taken to be
$e^{\pm i\pi s}$, depending on the election of the cut when
defining the complex power. As a consequence, the determinant
could get an undetermined phase \cite{ecz,woj}. We will come back
to this point later on, when performing the $s$-derivatives of
both zeta functions.

In order to do so, we must first perform an analytic extension of
both expressions, which are convergent for $\Re (s)$ big enough.
We first do it for $\zeta_1$ (a general discussion of the
procedure to be employed can be found, for instance, in
\cite{elilibro}).

We write equation (\ref{z1}) as a Mellin transform \beq
{\zeta}_1(s)=\frac{(1+(-1)^{-s})}{\Gamma(\frac{s}{2})}\int_0^{\infty}dt\,
t^{\frac{s}{2}-1}\sum_{\begin{array}{c}
  n=1 \\
  l=-\infty \\\end{array}}^{\infty}e^{-t\left[{\left(\frac{n\pi}{\alpha L}\right)}^2+
{\left((2l+1)\frac{\pi}{\alpha \beta}\right)}^2\right]}\,.\eeq

Now, using the definition of the Jacobi theta function \beq
\Theta_3 (z,x)=\sum_{l=-\infty}^{\infty}e^{-\pi x l^2} e^{2\pi z
l}\,,\label{th3}\eeq ${\zeta}_1$ can be rewritten as \beq \nn
{\zeta}_1
(s)&=&\frac{(1+(-1)^{-s})}{{(\sqrt{\pi}})^{s}\Gamma(\frac{s}{2})}\sum_{n=1}^{\infty}\int_0^{\infty}dt\,
t^{\frac{s}{2}-1}e^{-t\pi \left[\left(\frac{n}{\alpha
L}\right)^2+\left(\frac{1}{\alpha
\beta}\right)^2\right]}\times\\&&\Theta_3\left(\frac{-2t}{(\alpha
\beta)^2},\frac{4t}{(\alpha \beta)^2}\right)\,.\eeq

To proceed, we will use the inversion formula for the Jacobi
function \beq \Theta_3(z,x)=\frac{1}{\sqrt{x}}e^{(\frac{\pi
z^2}{x})}\Theta_3\left(\frac{z}{ix},\frac{1}{x}\right)\,,\eeq
together with the definition (\ref{th3}), thus getting \beq \nn
{\zeta}_1
(s)&=&\frac{(1+(-1)^{-s})\beta\alpha}{2{(\sqrt{\pi}})^{s}\Gamma(\frac{s}{2})}
\sum_{n=1}^{\infty}\int_0^{\infty}dt\, t^{\frac{s-1}{2}-1}e^{-t\pi
\left(\frac{n}{\alpha
L}\right)^2}\times\\&&\left[1+2\sum_{l=1}^{\infty}e^{i\pi
l}e^{-\frac{l^2 \pi \beta^2 \alpha^2}{4t}}\right]\,.\eeq

Now, the integrals can be performed to get \beq \nn {\zeta}_1
(s)&=&\frac{(1+(-1)^{-s})\beta}{2{\alpha}^{-s}{(\sqrt{\pi}})^{s}\Gamma(\frac{s}{2})}
\left[\Gamma\left(\frac{s-1}{2}\right)\frac{{\pi}^{\frac{1-s}{2}}}{
L^{1-s}}\zeta_R (s-1)+\right.\\&&\left.4\left(\frac{\beta L
}{2}\right)^{\frac{s-1}{2}}\sum_{n,l=1}^{\infty}(-1)^l
\left(\frac{l}{n}\right)^{\frac{s-1}{2}}K_{\frac{s-1}{2}}\left(\frac{nl\pi
\beta}{L}\right)\right]\,. \label{ext1}\eeq

This completes the analytic extension of $\zeta_1$. Note in
particular that, due to the behavior of the Bessel function
$K_{\nu}(z)$ for large $|z|$ \cite{rusa}, the series in the second term
between square brackets converges for all $s$.

We now extend $\zeta_2$. This is quite a simple task. In fact,
from the definition of the Hurwitz zeta function one can rewrite
equation (\ref{z2}) as follows \beq
\zeta_2(s)=\left(\frac{2\pi}{\beta
\alpha}\right)^{-s}(1+(-1)^{-s})\,\zeta_H
\left(s,\frac12\right)\,.\label{ext2}\eeq

We can now go to the evaluation of both contributions ($\Delta_1$
and $\Delta_2$) to the logarithm of the determinant. Before
actually doing so, note that $\zeta_1$ and $\zeta_2$ vanish for
$s=0$. This renders the ambiguity in defining $(-1)^{-s}$
irrelevant, even when taking the $s$-derivative. For the same
reason, all dependence on the unphysical parameter $\alpha$
disappears from such derivative. So, it is easy to see, from
equations (\ref{partfunc}), (\ref{del1}), (\ref{del2}),
(\ref{ext1}) and (\ref{ext2}), that \beq \nn \log{Z}&=&
-\frac{\beta}{2}\left[\frac{\pi}{6L}+\frac{4}{\beta}\sum_{n,l=1}^{\infty}\frac{(-1)^l}{l}\,e^{-\frac{nl\pi
\beta}{L}}\right]+\log{2}\\&=&-\frac{\beta}{2}\left[\frac{\pi}{6L}-\frac{4}{\beta}
\sum_{n=1}^{\infty}\log{(1+e^{-\frac{n\pi
\beta}{L}})}\right]+\log{2}\,. \eeq

From this result, the free energy can be obtained

\beq F=-\frac{1}{\beta}
\log{Z}=\frac{\pi}{12L}-\frac{2}{\beta}\sum_{n=1}^{\infty}\log{(1+e^{-\frac{n\pi
\beta}{L}})}-\frac{1}{\beta}\log{2}\,.\eeq

It is easy to see that, in the limit $\beta \rightarrow \infty$,
the right result for the Casimir energy at zero temperature
(equation (\ref{ec0})) is obtained.

The high temperature ($\beta \rightarrow 0$) behavior of the free
energy can be obtained by using the Euler-Maclaurin expansion,
thus getting \beq \nn F&=&-\frac{2L}{\pi
\beta^2}\int_0^{\infty}dx\,\log{(1+e^{-x})}+O(\frac{\beta^2}{L^3})\\&=&
-\frac{\pi L}{6\beta^2}+O(\frac{\beta^2}{L^3})\,.\label{hight}\eeq

Note the contribution from $\zeta_2$ cancels, in this limit, a
term linear in the temperature appearing in the Euler-Maclaurin
expansion.

\subsection{$\Th=\pi$}\label{Sect32}

In this case, there is only one contribution to the zeta function
and, consequently, to the partition function, i.e. \beq \zeta
(s)=(1+(-1)^{-s})\sum_{\begin{array}{c}
  n=0 \\
  l=-\infty \\
\end{array}}^{\infty}\left[{\left(\frac{\left(n+\frac12\right)\pi}{\alpha L}\right)}^2+
{\left((2l+1)\frac{\pi}{\alpha
\beta}\right)}^2\right]^{-\frac{s}{2}}\,.\label{z}\eeq

The steps leading to its analytic extension, and to the evaluation
of the partition function are essentially the same as in the
previous subsection. As a result, one obtains \beq \nn \zeta
(s)=\frac{(1+(-1)^{-s})\beta}{2{\alpha}^{-s}{(\sqrt{\pi}})^{s}\Gamma(\frac{s}{2})}
\left[\Gamma\left(\frac{s-1}{2}\right)\frac{{\pi}^{\frac{1-s}{2}}}{
L^{1-s}}\zeta_H \left(s-1,\frac12\right)+\right.\\
\left.4\left(\frac{\beta L
}{2}\right)^{\frac{s-1}{2}}\sum_{\begin{array}{c}
  n=0 \\
  l=1 \\
\end{array}}^{\infty}(-1)^l
\left(\frac{l}{n+\frac12}\right)^{\frac{s-1}{2}}
K_{\frac{s-1}{2}}\left(\frac{\left(n+\frac12\right)l\pi
\beta}{L}\right)\right]\,, \label{ext}\eeq and
\beq\log{Z}=\frac{\pi \beta}{24L}+2
\sum_{n=0}^{\infty}\log{(1+e^{-\frac{\left(n+\frac12\right)\pi
\beta}{L}})}\,. \eeq

Thus,

\beq F=-\frac{\pi }{24L}-\frac{2}{\beta}
\sum_{n=0}^{\infty}\log{(1+e^{-\frac{\left(n+\frac12\right)\pi
\beta}{L}})}\,. \eeq

In the $\beta \rightarrow \infty$ limit, one reobtains the Casimir
energy in section \ref{Sect1}. The high temperature limit
coincides with (\ref{hight}).

\section{Fermion number}\label{Sect4}

To evaluate the finite temperature fermion number, we will
introduce a small chemical potential (the meaning of ``small" will be specified, 
for each value of $\theta$, in the corresponding subsection) . Then, the finite temperature
fermion number will be calculated as \beq
N=\frac{1}{\beta}\frac{\partial\,\log{{\cal Z}}}{\partial
\mu} \,,\label{fnum}\eeq where ${\cal Z}$ is
the grand-canonical partition function. In the language of thermostatistics, this is the mean 
particle number of the Fermi-Dirac gas in the grand-canonical ensemble. The answer to the question 
posed in reference \cite{dunne} will be found from the value of this object at $\mu=0$ which includes, 
for each value of $\theta$,
both the zero-temperature fermion number (given in equation (\ref{nn}) and the paragraph following it) and its temperature-dependent part.

We will follow \cite{actor} in introducing the chemical potential
as an imaginary $A_0=i\mu$ in Euclidean space (or, equivalently, a
real $A_0$ in Minkowski space-time). The Dirac eigenvalue equation
then becomes \beq(\Dirac +i\gce \mu)\Psi=\omega \Psi\,,\eeq while
$\Psi$ again satisfies antiperiodic boundary conditions in the
``time" direction, and the ones given by (\ref{bc}) in the
``space" direction.

The chemical potential can be eliminated from the differential
equation through the transformation \beq \Psi=e^{-\mu
x_0}\Psi^{'}\,. \eeq

Thus, one gets for $\Psi ^{'}$ the same differential equation as
in the free case. Moreover, $\Psi^{'}$ satisfies the same
``spatial" boundary conditions but, in the ``time" direction, one
has \beq\Psi^{'}(\beta, x_1 )=-e^{\mu \beta}\Psi^{'}(0, x_1
)\,.\eeq

So, the effect of the chemical potential is to replace the
``temporal" eigenvalues in equation (\ref{lambda}) with \beq
\tilde{\lambda}=\lambda -i\mu= (2l+1)\frac{\pi}{\beta}-i\mu,\qquad
  l=-\infty,...,\infty\,.\eeq

\subsection{$\Th=0$}\label{Sect41}

As before, we must consider two contributions to $\log{{\cal Z}}$
\beq \Delta_1=\left. -\frac{d}{ds}\right\rfloor_{s=0}{\zeta}_1
(s)\,,\label{del1mu}\eeq where \beq {\zeta}_1
(s)=(1+(-1)^{-s})\sum_{\begin{array}{c}
  n=1 \\
  l=-\infty \\
\end{array}}^{\infty}\left[{\left(\frac{n\pi}{\alpha L}\right)}^2+
{\left((2l+1)\frac{\pi}{\alpha
\beta}-i\frac{\mu}{\alpha}\right)}^2\right]^{-\frac{s}{2}}\,,\label{z1mu}\eeq
and \beq \Delta_2=\left. -\frac{d}{ds}\right\rfloor_{s=0}{\zeta}_2
(s)\,,\label{del2mu}\eeq where \beq  {\zeta}_2 (s)=\sum_{l=-\infty
}^{\infty}\left[ (2l+1)\frac{\pi}{\alpha
\beta}-i\frac{\mu}{\alpha}\right]^{-s}\,.\label{z2mu}\eeq

In order to perform the analytic extension of $\zeta_1$, we will
proceed much the same way as in the $\mu=0$ case. However, in
order to properly write (\ref{z1mu}) in terms of its Mellin
transform, and freely interchange integrals and double series, we
will take $\mu$ to satisfy $|\mu|<\frac{\pi}{L}$. Thus, we can
write \beq
{\zeta}_1(s)=\frac{(1+(-1)^{-s})}{\Gamma(\frac{s}{2})}\int_0^{\infty}dt\,
t^{\frac{s}{2}-1}\sum_{\begin{array}{c}
  n=1 \\
  l=-\infty \\\end{array}}^{\infty}e^{-t\left[{\left(\frac{n\pi}{\alpha L}\right)}^2+
{\left((2l+1)\frac{\pi}{\alpha
\beta}-i\frac{\mu}{\alpha}\right)}^2\right]}\,.\eeq

This can also be written as \beq \nn {\zeta}_1
(s)&=&\frac{(1+(-1)^{-s})}{{(\sqrt{\pi}})^{s}\Gamma(\frac{s}{2})}\sum_{n=1}^{\infty}\int_0^{\infty}dt\,
t^{\frac{s}{2}-1}e^{-t\pi \left[\left(\frac{n}{\alpha
L}\right)^2+\left(\frac{1}{\alpha \beta}-\frac{i\mu}{\alpha
\pi}\right)^2\right]}\times\\&&\Theta_3\left(\frac{-2t}{\alpha
\beta}\left(\frac{1}{\alpha \beta}-\frac{i\mu}{\alpha\pi
}\right),\frac{4t}{(\alpha \beta)^2}\right)\,.\eeq

From here on, the same steps as in the previous section can be
followed, to obtain \beq \nn {\zeta}_1
(s)&&=\frac{(1+(-1)^{-s})\beta}{2{\alpha}^{-s}{(\sqrt{\pi}})^{s}\Gamma(\frac{s}{2})}
\left[\Gamma\left(\frac{s-1}{2}\right)\frac{{\pi}^{\frac{1-s}{2}}}{
L^{1-s}}\zeta_R (s-1)+\right.\\&&\left.4\left(\frac{\beta L
}{2}\right)^{\frac{s-1}{2}}\sum_{n,l=1}^{\infty}(-1)^l
\left(\frac{l}{n}\right)^{\frac{s-1}{2}}\cosh{(\mu \beta
l)}K_{\frac{s-1}{2}}\left(\frac{nl\pi \beta}{L}\right)\right]\,.
\label{ext1mu}\eeq

Note the double sum in the last term between square brackets is
convergent in the range of $\mu$ considered. So, again,
$\zeta_1(0)$=0

The analytic extension of $\zeta_2$ requires, in this case, a more
careful treatment than in the case $\mu$=0. In fact, \beq \nn
{\zeta}_2 (s)&=&\sum_{l=-\infty }^{\infty}\left[
(2l+1)\frac{\pi}{\alpha
\beta}-i\frac{\mu}{\alpha}\right]^{-s}\\
\nn &=&\left(\frac{2\pi}{\alpha\beta}\right)^{-s}\left[\sum_{l=0
}^{\infty}\left[
(l+\frac12)-i\frac{\mu\beta}{2\pi}\right]^{-s}+\sum_{l=0
}^{\infty}\left[
-(l+\frac12)-i\frac{\mu\beta}{2\pi}\right]^{-s}\right]\\&=&
 \left(\frac{2\pi}{\alpha\beta
}\right)^{-s}\left[ \zeta_H \left(s,\frac12-\frac{i\mu
\beta}{2\pi}\right)+\sum_{l=0 }^{\infty}\left[
-(l+\frac12)-i\frac{\mu\beta}{2\pi}\right]^{-s}\right]\,.\eeq

Now, in order to write the second term as a Hurwitz zeta, we must
relate the eigenvalues with negative real part to those with
positive one without, in so doing, going through zeros in the
argument of the power. Otherwise stated, we must select a cut in
the complex $\omega$ plane \cite{ecz}. This requirement determines
a definite value of $(-1)^{-s}$, i.e., $(-1)^{-s}=e^{i\pi sign
(\mu)s}$. Taking this into account, we finally have \beq
\zeta_2(s)=\left(\frac{2\pi}{\beta
\alpha}\right)^{-s}\left[\zeta_H \left(s,\frac12-\frac{i\mu
\beta}{2\pi}\right)+e^{i\pi sign(\mu)s}\zeta_H
\left(s,\frac12+\frac{i\mu \beta}{2\pi}\right)\right]
\label{ext2mu}\,.\eeq

From (\ref{ext1mu}) and (\ref{ext2mu}) both contributions to
$\log{{\cal Z}}$ can be obtained. They are given by \beq
\Delta_1=-\frac{\beta
\pi}{12L}+\sum_{n=1}^{\infty}\log{(1+e^{-\frac{2n\pi
\beta}{L}}+2\cosh{(\mu \beta)}e^{-\frac{n\pi \beta}{L}})}\,\eeq
and \beq\nn \Delta_2 &=&-\left[\zeta_H^{\prime}\left(0, \frac12
-\frac{i\mu \beta}{2\pi}\right)+\zeta_H^{\prime}\left(0, \frac12
+\frac{i\mu \beta}{2\pi}\right)+i\pi sign(\mu)\zeta_H \left(0,
\frac12 +\frac{i\mu \beta}{2\pi}\right)\right]\\
&=&\log{2}+\log{\cosh{\left(\frac{\mu
\beta}{2}\right)}}-\frac{|\mu|\beta}{2}\,.\eeq

Both expressions can be seen to reduce to the corresponding ones
in the previous section when $\mu=0$.

Putting both pieces together, we finally have \beq\nn\log{{\cal
Z}}&=&-\frac{\beta
\pi}{12L}+\sum_{n=1}^{\infty}\log{(1+e^{-\frac{2n\pi
\beta}{L}}+2\cosh{(\mu \beta)}e^{-\frac{n\pi \beta}{L}})}\\
&+&\log{2}+\log{\cosh{\left(\frac{\mu
\beta}{2}\right)}}-\frac{|\mu|\beta}{2}\,.\eeq

From this result, we can evaluate the free energy \beq\nn
F&=&\frac{
\pi}{12L}-\frac{1}{\beta}\left[\sum_{n=1}^{\infty}\log{(1+e^{-\frac{2n\pi
\beta}{L}}+2\cosh{(\mu \beta)}e^{-\frac{n\pi \beta}{L}})}\right.\\
&+&\left.\log{2}+\log{\cosh{\left(\frac{\mu
\beta}{2}\right)}}-\frac{|\mu|\beta}{2}\right]\,.\eeq

It is easy to see that its $\beta \rightarrow \infty$ limit is
$\frac{ \pi}{12L}$, independently of the value of $\mu$. This is
consistent with the fact that the chemical potential has been
introduced as a purely imaginary $A_0$ gauge potential in
Euclidean space. This corresponds to a real $A_0$ potential in
Minkowski space-time. Now, as is well known, this last can be
eliminated at zero temperature through a gauge transformation.

The high-temperature limit of the free energy is $-\frac{\pi
L}{6\beta^2}-\frac{L\mu^2}{2\pi}+\frac{|\mu|}{2}$.

According to (\ref{fnum}), the finite-temperature fermion number
is given by \beq \nn
N&=&\left\{\sum_{n=1}^{\infty}\left[\frac{e^{-\frac{n\pi
\beta}{L}+\mu \beta}}{1+e^{-\frac{n\pi \beta}{L}+\mu
\beta}}-\frac{e^{-\frac{n\pi \beta}{L}-\mu
\beta}}{1+e^{-\frac{n\pi \beta}{L}-\mu
\beta}}\right]\right.\\&+&\left.
\frac12\tanh\left(\frac{\mu \beta}{2}\right)-\frac12
sign(\mu)\right\}\,.\label{numero} \eeq
In particular, for $\mu=0$, it is clearly undefined. Note that this discontinuous character of
the fermion number comes from zero "spatial" eigenvalues or,
equivalently, from the spectral asymmetry. More precisely, it
originates from the phase of the determinant. We will comment on
the interpretation of this result in the last section of the
paper.

\subsection{$\Th=\pi$}\label{Sect42}

Since, for this value of $\Th$, only the symmetric part of the
spectrum contributes, we must consider the zeta function \beq
\zeta (s)=(1+(-1)^{-s})\sum_{\begin{array}{c}
  n=0 \\
  l=-\infty \\
\end{array}}^{\infty}\left[{\left(\frac{\left(n+\frac12\right)\pi}{\alpha L}\right)}^2+
{\left((2l+1)\frac{\pi}{\alpha
\beta}-i\frac{\mu}{\alpha}\right)}^2\right]^{-\frac{s}{2}}\,,\label{zmu}\eeq
whose analytic extension can be obtained with the same method as
before, and is given (for $|\mu|<\frac{\pi}{2L}$) by \beq \nn
\zeta
(s)=\frac{(1+(-1)^{-s})\beta}{2{\alpha}^{-s}{(\sqrt{\pi}})^{s}\Gamma(\frac{s}{2})}
\left[\Gamma\left(\frac{s-1}{2}\right)\frac{{\pi}^{\frac{1-s}{2}}}{
L^{1-s}}\zeta_H \left(s-1,\frac12\right)+\right.\\
\left.4\left(\frac{ \beta L
}{2}\right)^{\frac{s-1}{2}}\sum_{\begin{array}{c}
  n=0 \\
  l=1 \\
\end{array}}^{\infty}(-1)^l
\left(\frac{l}{n+\frac12}\right)^{\frac{s-1}{2}} \cosh{(\mu \beta
l)} K_{\frac{s-1}{2}}\left(\frac{\left(n+\frac12\right)l\pi
\beta}{L}\right)\right]\,. \label{extmu}\eeq

The grand-canonical partition function, obtained by evaluating its
$s$-derivative at $s=0$ with reversed sign, gives us\beq\log{{\cal
Z}}=\frac{\beta
\pi}{24L}+\sum_{n=0}^{\infty}\log{(1+e^{-\frac{2\left(n+\frac12\right)\pi
\beta}{L}}+2\cosh{(\mu \beta)}e^{-\frac{\left(n+\frac12\right)\pi
\beta}{L}})}\,\eeq

In this case one has, for the free energy, \beq F=-\frac{
\pi}{24L}-\frac{1}{\beta}\sum_{n=0}^{\infty}\log{(1+e^{-\frac{2\left(n+\frac12\right)\pi
\beta}{L}}+2\cosh{(\mu \beta)}e^{-\frac{\left(n+\frac12\right)\pi
\beta}{L}})}\,.\eeq

Its low-temperature limit is $-\frac{ \pi}{24L}$ which, again, is
$\mu$-independent. The high-temperature limit is $-\frac{\pi
L}{6\beta^2}-\frac{L\mu^2}{2\pi}$.

The fermion number is given by \beq
N&=&\sum_{n=0}^{\infty}\left[\frac{e^{-\frac{\left(n+\frac12\right)\pi
\beta}{L}+\mu \beta}}{1+e^{-\frac{\left(n+\frac12\right)\pi
\beta}{L}+\mu \beta}}-\frac{e^{-\frac{\left(n+\frac12\right)\pi
\beta}{L}-\mu \beta}}{1+e^{-\frac{\left(n+\frac12\right)\pi
\beta}{L}-\mu \beta}}\right]\,,\label{pi}\eeq
which vanishes for $\mu=0$.

\section{Discussion of the results}\label{Sect5}

The result for the finite-temperature fermion number in the case
$\Th=\pi$ (equation (\ref{pi}) has a clear interpretation: no
fermion number is created when rising the temperature, while
keeping $\mu=0$. Moreover, in this case, the low temperature limit
of the fermion number also vanishes for $\mu\neq 0$.

Now, in the case $\Th=0$, the result in (\ref{numero}) requires a
careful analysis. From the point of view of field theory, the
right answer must be taken as one of the two possible limits. In
fact, at zero temperature, a zero mode of the Hamiltonian appears
in Minkowski space for $\mu=0$ and, as already discussed (see
section \ref{Sect1}) there will be two nonequivalent vacuum
states, with $N=\pm\frac12$. Once the theory is quantized around
one of these vacuum states, no extra fermion number ($U(1)$
charge) will arise at one loop, when compactifying the ``temporal"
coordinate with an antiperiodic twist. This is, for $\theta=0$,
the answer to the question posed in \cite{dunne}. However, the
low-temperature limit of the fermion number can be seen to vanish
if $\mu$ is kept different from zero, due to thermal averaging
over the degenerate ground states.

It is interesting to note that our result for the partition
function in this (Ramond) case doesn't coincide with the analytic
extension of the one given, for instance, in Chapter 10 of
reference \cite{polchinski} (see also \cite{lugo}), where the contribution from the phase
of the determinant doesn't appear. Were one to disregard this
term, the low-temperature limit of the fermion number would vanish
for $\mu=0$ (instead of picking one of the two possible vacuum
values), while it would be $\frac12 sign (\mu)$ for $\mu\neq 0$.

$\phantom{aa}$\\
{\bf Acknowledgements:} We thank Horacio Falomir and, especially,
Adri\'an Lugo for useful discussions and suggestions.

This work was partially supported by CONICET (Grant 0459) and UNLP
(Grant X298).

\end{document}